\documentclass[fleqn,usenatbib]{mnras}

\usepackage{mathptmx}
\usepackage[T1]{fontenc}

\DeclareRobustCommand{\VAN}[3]{#2}
\let\VANthebibliography\thebibliography
\def\thebibliography{\DeclareRobustCommand{\VAN}[3]{##3}\VANthebibliography}

\usepackage{graphicx}	
\usepackage{amsmath}	
\usepackage{amssymb}	
\usepackage{subcaption}
\usepackage{multicol}
\usepackage{verbatim}
\usepackage{multirow}
\usepackage{booktabs} 
\usepackage{tablefootnote} 

\newcommand{\src}{XTE~J$1701-462$\,}

\title[X-ray polarization from XTE J1701$-$462]{Detection of significant X-ray polarization from transient NS-LMXB XTE J1701$-$462 with IXPE and its implication on the coronal geometry}

\author[Jayasurya et al.]{
	Kiran M. Jayasurya$^{1}$\thanks{E-mail: kiranmj@ursc.gov.in}, Vivek K. Agrawal$^1$ and Rwitika Chatterjee$^1$ \\
	$^1$Space Astronomy Group, ISITE Campus, U. R. Rao Satellite Centre, Bangalore, 560037, India 
}

\date{Accepted XXX. Received YYY; in original form ZZZ}

\pubyear{2023}

\begin{document}
\label{firstpage}
\pagerange{\pageref{firstpage}--\pageref{lastpage}}
\maketitle

\begin{abstract}
In this paper, we performed a spectro-polarimetric analysis of the transient NS-LMXB \src using \textit{IXPE}, \textit{NICER} and \textit{NuSTAR} data during its 2022 outburst. We report significant detection of energy-dependent polarization in the X-ray signal from the source on 2022~September~29 in the $2-4$~keV, $4-8$~keV and $2-8$~keV energy bands with a  polarization degree of 3.9 $\pm$ 0.3\%, 5.5 $\pm$ 0.6\% and 4.5 $\pm$ 0.4\%, respectively. The polarization angle in the overall $2-8$~keV band was $\sim 143^{\circ}\pm 2^{\circ}$. The spectra were modelled using a combination of thermal emission from an accretion disc, Comptonized emission from a hot electron plasma (or corona) and a Gaussian line. From spectro-polarimetric analysis, the polarization degree due to the disc emission had an upper limit of $\sim$ 11.5\%, and that of the Comptonized emission was constrained at 7.7$\pm$2.5\% (at the 3$\sigma$ level). The results suggest that the Comptonized component probably originates from a spreading-layer/boundary-layer above the neutron star surface. IXPE observation of the source on 2022~October~8 does not show significant polarization which can be attributed to a weakening of the coronal Comptonized emission. The implication of these results are discussed.
\end{abstract}
%
%
%
%
\section{Introduction}
\quad
Low-mass X-ray binaries (LMXBs) consisting of a low-magnetic field  neutron star accreting material from a low-mass companion star show intensity and spectral variations on time scales ranging from hours to months, and are divided into two sub-groups, Z-sources and atoll sources. The classification is  based on their joint spectral and temporal nature and the shape of the path that they trace out in the Colour-Colour diagram (CCD) and hardness intensity diagram \citep[HID,][]{has1989}. The Z-sources trace out a  `Z'-shaped track having three branches: horizontal branch (HB), normal branch (NB) and flaring branch (FB). The Z-sources are further classified into two sub-categories based on the extent of their HB and FB. The Cyg-like
Z-sources have prominent HB and weak FB, and vice-versa for the Sco-like Z-sources.

\src is a neutron star low-mass X-ray binary (NS-LMXB) discovered with the \textit{All Sky Monitor (ASM) of the \textit{Rossi X-ray Timing Explorer}} (\textit{RXTE}) in January~2006 \citep{remi2006} and it is the only known source till date that shows both Z-like and Atoll-like behaviour. The first 10 weeks of observations suggested that it was a transient Z-source, the first of its kind \citep{homan2007}. During the initial high intensity phase, the source displayed Cyg-like behaviour and then evolved into a Sco-like Z-source as the count rate decreased \citep{lin2009}. During the decay phase of the $\sim 600$ days outburst, the source displayed atoll-like behaviour. MAXI/GSC observations has revealed that the source entered  a new outburst phase on 2022 September 6 \citep{iwa2022}.

In general, a combination of a thermal component and a Comptonized emission component is used in modeling the X-ray spectra of Z and atoll sources  \citep{disal2000,disal2001,agra2003,agra2009,agra2023,tara2008,pirai2000,pira2007}.
The thermal component is either described by a multi-colour disc (MCD) component from standard thin disc \citep{mitsu1984} or a single-temperature blackbody (BB) originating from the boundary-layer/spreading-layer (BL/SL).The Comptonized emission is produced by scattering of soft seed photons in a hot electron plasma. The seed photons are supplied by either relatively cold accretion disc or the neutron star surface.
\par \cite{lin2009} fitted the X-ray spectrum of  XTE 1701-462 with a combination of two thermal components (BB+MCD) and a constrained broken power-law (CBPL) with  break energy fixed at 20 keV. A combination of a Comptonized component (\texttt{compTT} in XSPEC) and disc emission (MCD) has also been used to describe the X-ray spectrum of the source \citep{wang2014}. Z-sources are generally found in the soft spectral state (SS). In this state, the Comptonized component $kT_{e}$ is found to be in the range of 2 to 5 keV  and the spectrum  shows small but systematic variations along the complete Z-track \citep{disal2000,disal2001,disal2002}.

In order to constrain the radiative processes at play, the accretion flow geometry and orbital inclination in LMXBs, polarization data in the X-ray band is extremely useful. Detailed polarimetric simulations  have been performed for some of the basic configurations (slab, wedge and shell type) of corona in compact objects \citep{gnar2022,sch2010}. Some of the LMXBs are a prime target of the \textit{Imaging X-ray Polarimetry Explorer} (\textit{IXPE}), a polarimetric mission  launched
on 2021 December 9 \citep{weiss2022}. These observations are providing a better view of the physical processes and geometry of the putative corona in these systems. 
Recently, significant polarization has been detected in LMXBs like Cyg X-2 \citep{fari2023} and GX 9+9 \citep{rwicha2023,urs2023} with IXPE, and Sco X-1 with \textit{PolarLight} \citep{long2022}. In the case of Sco X-1 and Cyg X-2, the PA was found to be aligned with the normal to the disc plane suggesting hot electron plasma (Compton corona) situated above the transition or spreading layer. No significant polarization was detected in the atoll source GS 1826$-$238  \citep{capi2023} with IXPE, constraining the system inclination and coronal geometry. 

In this paper, we report the detection of polarized
X-ray signal from \src using \textit{IXPE} data. We discuss the implication of our results and suggest
a possible accretion flow and corona geometry in the source.

\section{Observations and Data Analysis}
\label{sec:obs}
\subsection{\textit{IXPE}}
\quad

\textit{IXPE} is a Gas Pixel Detector (GPD)-based soft X-ray imaging polarimeter \citep{Soffitta2021,weiss2022}, comprising of a 4~m focal length Mirror Module Assembly focusing X-rays onto three polarization-sensitive detector units (DUs) in the $2-8$~keV energy range. \textit{IXPE} observed \src on 2022~September~29 (ObsID 01250601; hereafter referred to as Epoch~1) and again on 2022~October~8 (ObsID 01250701; hereafter referred to as Epoch~2) for 46.2 ks and 46.4 ks of net exposure times respectively.






	
	


The processed \textit{IXPE} Level-2 data\footnote{Publicly available on the HEASARC Data Archive} was analyzed using \texttt{IXPEOBSSIM} v30.0.0 \citep{baldini}. The source region was defined as a 60" circle and the background region as an annulus of 180" inner and 240" outer radii respectively, centered at the source's centroid intensity. The source and background event lists were extracted using the \texttt{XPSELECT} task, following which the \texttt{PCUBE} algorithm of \texttt{XPBIN} task was used to generate the polarization cubes. The \texttt{PHA1}, \texttt{PHA1Q} and \texttt{PHA1U} algorithms were used to generate the Stokes I, Q and U spectra respectively.

The polarization degree (PD) and polarization angle (PA) were determined using the model-independent \texttt{PCUBE} algorithm \citep{KISLAT201545} for the energy bands $2-4$~keV, $4-8$~keV and $2-8$~keV. The algorithm assumes that the PD and PA are independent, but they are not so in reality, and hence the uncertainties are more appropriately represented by the contours of their joint measurement.

The results of our polarimetric analysis for both IXPE Epochs are described in Section~\ref{sec:sppol}. 
We also performed a spectro-polarimetric model-dependent fit \citep{stro2017} using \textit{XSPEC} and the latest \textit{IXPE} response files (v12).

\subsection{\textit{NICER} and \textit{NuSTAR}}
\quad
The \textit{Neutron Star Interior Composition ExploreR  (NICER)} observed \src on 2022~September~29 (ObsID 5203390122) for 2784 s of net exposure time. The cleaned event files were extracted using \texttt{nicerl2} task which performs standard calibration and screening of the unfiltered data. The source and background spectra (in the $0.2-12$~keV energy band) were generated using \texttt{nicerl3-spect} task. The tasks were performed using \textit{NICERDAS} software v10 distributed with \textit{HEASOFT} v6.31.1\footnote{http://heasarc.gsfc.nasa.gov/ftools} and the latest CALDB.



The source was also observed by the \textit{Nuclear Spectroscopic Telescope Array (\textit{NuSTAR})} on 2022~October~8 (ObsID 90801325002) for a net exposure time of 12.2 ks. The unfiltered event files were calibrated and screened with the \texttt{nupipeline} task of \textit{NuSTARDAS} (v1.9.7) using the latest CALDB files. The spectra were generated using the \texttt{nuproducts} task.

The \textit{NICER} and \textit{NuSTAR} quasi-simultaneous observations were considered for the spectro-polarimetric analysis of the IXPE Epoch 1 and Epoch 2 data respectively.
\begin{figure}
	\centering
	\begin{subfigure}[t]{0.34\textwidth}
		\includegraphics[width=\columnwidth]{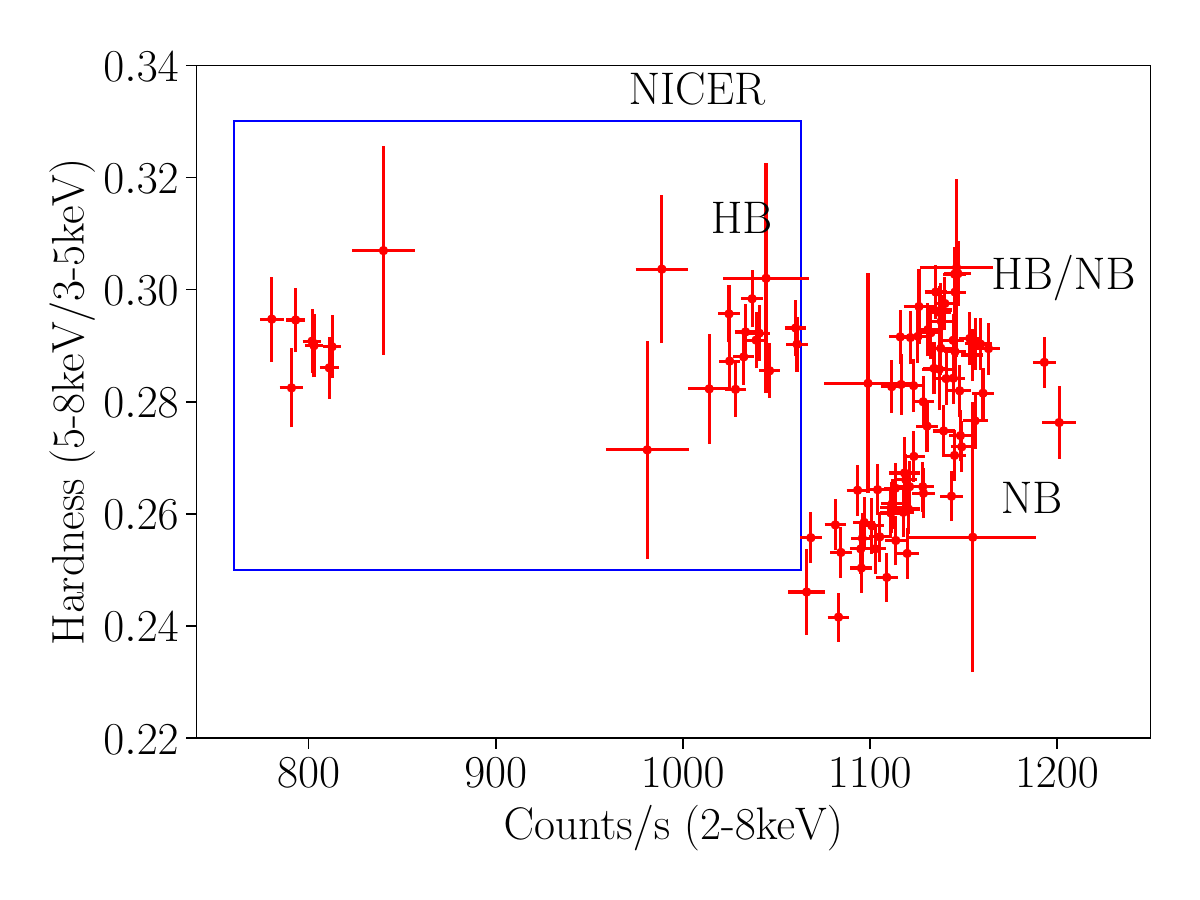}
	\end{subfigure}
	\begin{subfigure}[t]{0.34\textwidth}
		\includegraphics[width=\columnwidth]{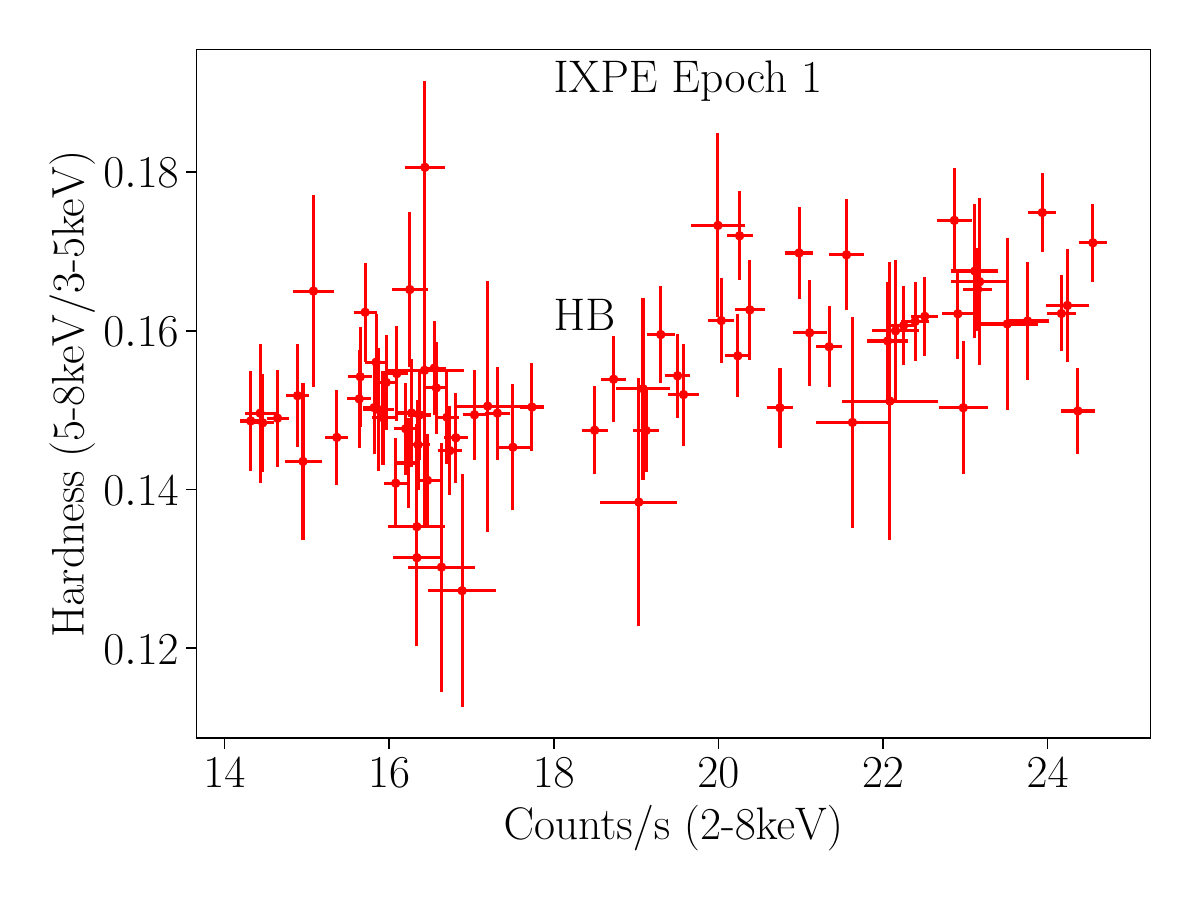}
	\end{subfigure}
	\caption{HIDs for \textit{NICER} observation (top) and IXPE Epoch 1 (bottom, using the DU1). The blue box represents the region from which the \textit{NICER} spectrum was extracted for analysis.}
	\label{fig:hide1}
\end{figure}

Since the NICER observation is close to Epoch~1, light curves in $3-5$~keV, $5-8$~keV and $2-8$~keV bands were generated to create a HID of the source (Figure \ref{fig:hide1}). The hardness was considered as the ratio of the counts in the $5-8$~keV band to the $3-5$~keV band, and the intensity was taken as the count rate in the $2-8$~keV band. An HID was also created for the corresponding \textit{IXPE} data using the DU1 in the same energy bands. The IXPE HID showed that the source was in the extended HB. Hence, we extracted the \textit{NICER} spectrum in the HB (selection region is shown as a blue box in Figure \ref{fig:hide1}). 

During Epoch~2, part of the \textit{NuSTAR} data overlapped with the \textit{IXPE} observation period. A common good time interval (GTI) file was created by merging the GTI files of the \textit{NuSTAR} and \textit{IXPE} data using the \texttt{mgtime} task of \textit{FTOOLS}. This file was then used to extract the \textit{NuSTAR} spectra (for both FPMA and FPMB) for the spectral analysis. 

Both \textit{NICER} and \textit{NuSTAR} spectra were re-binned to have minimum 25 counts per energy bin and the fitting was done in the $1.8-12$~keV and $3-78$~keV energy bands respectively.

\section{Results}
\label{sec:results}
\subsection{Spectral Properties}
\label{sec:specresults} 
\quad

\begin{figure}
	\centering
	\vspace{-1cm}
	\quad \hspace*{0.2cm}\includegraphics[width=1.35\columnwidth,trim=3.5cm 3.5cm 0cm 4cm, clip=true]{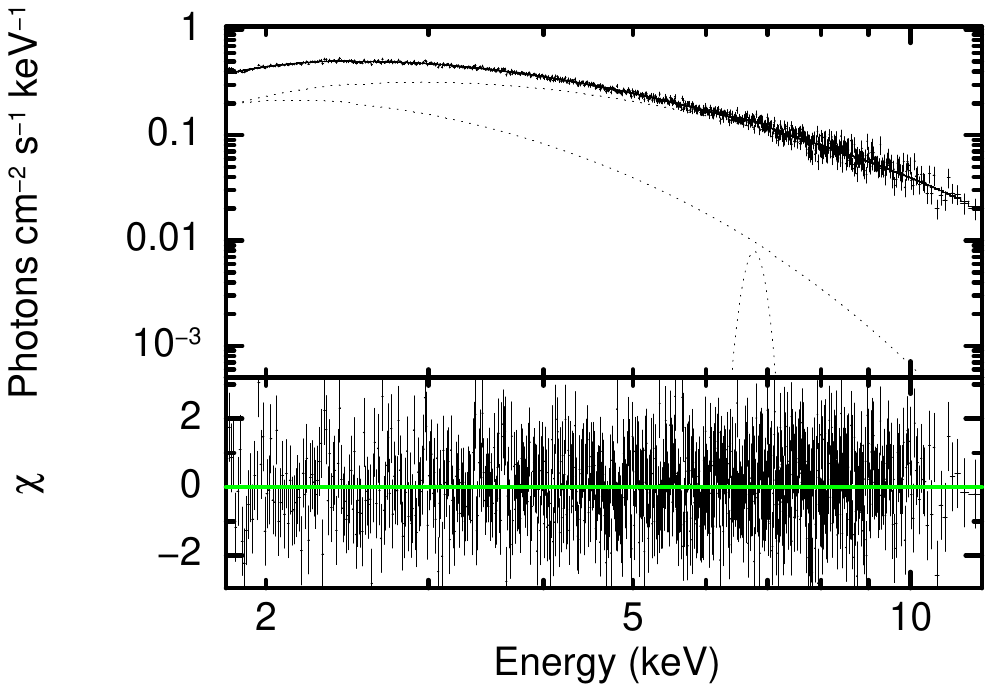}
	\caption{(Top panel) Unfolded \textit{NICER} spectrum (corresponding to the region marked in Figure~\ref{fig:hide1}) of the source for Epoch 1, fitted with model 1. (Bottom panel) Residuals of the fit in units of $\sigma$.}
	\label{fig:combspecfit}
\end{figure}

The \textit{NICER} spectrum (corresponding to the region marked in Figure~\ref{fig:hide1}) was fitted in XSPEC with three different models, which are widely used to describe NS-LMXB spectra: \texttt{tbabs*(diskbb+nthComp+gaussian)} (hereafter, model~1), \texttt{tbabs*(bbodyrad+diskbb+gaussian)} (hereafter, model~2) and \texttt{tbabs*(bbodyrad+nthComp+gaussian)} (hereafter, model~3).

In these models, \texttt{diskbb} component represents emission from a standard accretion disc  \citep{mitsu1984}, \texttt{nthComp} describes the Comptonized emission from a hot electron plasma \citep[see][]{Zd1996} and \texttt{bbodyrad} represents the single temperature blackbody emission from the neutron star surface. Model 1 represents a scenario where the soft emission comes from an accretion disc, the hard component from a corona which can have either spherical or slab-type geometry, and the seed photons for Comptonization originate from the neutron star surface. In model 2, the hotter component comes from the neutron star surface/BL, and cooler component originates from the accretion disc. Model 3 represents a scenario where the soft component comes from the neutron star surface/BL and the hard component from a corona having slab (or wedge) type geometry. The seed photons in this case are provided by the accretion disc. In all three scenarios, the iron emission line is modelled by a gaussian component and is believed to be produced by the reflection of coronal (in model 1 or 3) or BB (in model 2) radiation over the inner accretion disc. This reflection feature is usually associated with a broad continuum, an iron line and an absorption edge which are smeared due to relativistic effects.

All three models produced acceptable fits to the spectra and the best-fit parameters are tabulated in Table~\ref{tab:specfittab}. 
The  \textit{NICER}  spectrum fitted with model 1 is also shown in Figure~\ref{fig:combspecfit}.
We also fit the \textit{NuSTAR} spectra with these models multiplied by a \texttt{smedge} \citep[smeared edge,][]{ebisawa1991,ebi1994} component,
due to the presence of a weak reflection feature, resulting in better reduced chi-square values. 
 The best-fit parameters are shown in Table \ref{tab:specfittabobs2}.
\begin{table}
	\centering
	\caption{Best fit parameters for the \textit{NICER} spectrum (for the region marked in Figure~\ref{fig:hide1}) for Epoch 1. $N_{H}$: Hydrogen column density; $kT_{bb}$: blackbody temperature; $kT_{in}$: inner disc temperature; $\Gamma_{nth}$: photon index; $kT_{s,nth}$: seed photon temperature for \texttt{nthcomp}; $kT_{e}$: electron plasma temperature; $\tau$ is the optical depth of the corona (assuming spherical geometry). $N_{bb}$, $N_{diskbb}$ and $N_{nth}$
		are the normalizations for the \texttt{bbodyrad}, \texttt{diskbb} and \texttt{nthcomp} components respectively. $E_{l}$,  $\sigma_{l}$, and $EW$  are the center, width, and equivalent width of the Gaussian line respectively.  $F_{bb}$, $F_{diskbb}$ and $F_{nth}$ are the unabsorbed fluxes in the $3-10$~keV range for the \texttt{bbodyrad}, \texttt{diskbb} and \texttt{nthcomp} components respectively, in units of 10$^{-9}$ ergs cm$^{-2}$ s$^{-1}$. Uncertainties are quoted at the 1$\sigma$ confidence range.}
	\label{tab:specfittab}
	\resizebox{0.85\columnwidth}{!}{
		\begin{tabular}{cccc} 
			\hline
			Parameter & Model 1 & Model 2 & Model 3 \\ 
			\hline
			$N_{H}$ (10$^{22}$ cm$^{-2}$) & 1.7$^{*}$ & 1.9$^{*}$ & 1.9$^{*}$\\
			$kT_{bb}$ (keV) & - &  2.4$_{-0.3}^{+0.5}$ &1.0$\pm$0.1\\
			$N_{bb}$ (keV) & - & 12$_{-7}^{+11}$  &  304$_{-34}^{+51}$\\
			$kT_{in}$ (keV) & 0.89$^{*}$ & 1.7$\pm$0.1 & -\\
			$N_{diskbb}$ & 333$_{-33}^{+25}$ & 93$_{-12}^{+11}$ &- \\
			$\Gamma_{nth}$ & 2.0$_{-0.2}^{+0.3}$ &  - & 1.7$_{-0.2}^{+0.3}$\\
			$kT_{e}$ (keV)& 2.3$_{-0.2}^{+0.4}$ & -  & 2.2$_{-0.1}^{+0.2}$ \\
			$kT_{s,nth}^{\dagger}$ (keV)& 0.91$_{-0.08}^{+0.07}$ & -  & 0.65$^{*}$\\
			$N_{nth}$ & 0.55$_{-0.04}^{+0.06}$ & -  & 1.8$\pm$0.1  \\	
			$E_{l}$ (keV) & 6.8$\pm$0.1 & 6.8$\pm$0.1 & 6.8$\pm$0.1\\	
			$\sigma_{l}$ (keV) & 0.16$_{-0.07}^{+0.15}$ & 0.15$_{-0.06}^{+0.14}$  & 0.16$_{-0.06}^{+0.16}$\\	
			$EW$ (eV) & 24 & 23 & 24 \\
			$\tau$ & 11 $\pm$ 3 & - & 7.7 $\pm$ 0.9\\
			$F_{diskbb}$ & 0.9 & 8.9 & -\\
			$F_{bb}$ & - & 2.6 & 2.0 \\
			$F_{nth}$ & 10 & - & 9.5 \\					        
			$\chi^{2}$/DOF & 778/762 & 782/763  & 780/762\\	        		
			\hline
			\multicolumn{4}{l}{* Frozen at best fit value.
			}\\
		\multicolumn{4}{l}{$\dagger$ Black body seed photons for model 1, disc seed photons for model 3}
		
	\end{tabular}}
\end{table}

\begin{table}
	\centering
	\caption{Best fit parameters for the \textit{NuSTAR} spectrum for Epoch 2. Parameter definitions, units and uncertainty levels are same as in Table \ref{tab:specfittab}. $E_{smedge}$ and $f_{max,smedge}$ are the threshold energy and maximum absorption factor at the threshold energy, respectively, for the smeared edge component.}
	\label{tab:specfittabobs2}
	\resizebox{0.85\columnwidth}{!}{
	\begin{tabular}{cccc} 
		\hline
		Parameter & Model 1 & Model 2 & Model 3 \\ 
		\hline
		$N_{H}$ (10$^{22}$ cm$^{-2}$) & 2.0$^{*}$ & 1.9$^{*}$ & 2.1$^{*}$\\
		$E_{smedge}$ (keV) & 8.8$\pm$0.1 & 8.8$\pm$0.1 & 8.8$_{-0.2}^{+0.1}$\\
		$f_{max,smedge}$ & 6.8$\pm$0.8 & 7.1$\pm$0.9 & 3.1$\pm$0.7\\
		$kT_{bb}$ (keV) & - &  2.5$\pm$0.1  & 1.2$\pm$0.1\\
		$N_{bb}$ (keV) & - & 2.1$_{-0.3}^{+0.4}$  &  327$\pm$10\\
		$kT_{in}$ (keV) & 1.4$\pm$0.1 & 1.6$\pm$0.1 & -\\
		$N_{diskbb}$ & 225$_{-49}^{+40}$ & 150$\pm$3 &- \\
		$\Gamma_{nth}$ & 4.6$\pm$0.1 &  - & 2.5$\pm$0.1\\
		$kT_{e}$ (keV)& 4.5$^{*}$ & -  & 2.7$\pm$0.1 \\
		$kT_{s,nth}$ (keV)& 1.4$\pm$0.1 & -  & 0.33$^{*}$\\
		$N_{nth}$ & 0.10$_{-0.04}^{+0.03}$ & -  & 6.1$\pm$0.3  \\						
		$E_{l}$ (keV) & 6.5$\pm$0.1& 6.5$\pm$0.1 & 6.5$\pm$0.1\\				
		$\sigma_{l}$ (keV) & 0.25$\pm$0.05 & 0.25$_{-0.04}^{+0.05}$  & 0.18$\pm$0.04\\	
		$EW$ (eV) & 44 & 43 & 32 \\
		$\tau$ & 2.6 $\pm$ 0.3 & - & 7.5 $\pm$ 0.4\\
		$F_{diskbb}$ & 6.8 & 9.9 & -\\
		$F_{bb}$ & - & 0.5 & 5.4 \\
		$F_{nth}$ & 3.6 & - & 5.2 \\					        
		$\chi^{2}$/DOF & 793/725 & 789/726  & 776/725\\	
		\hline
		\multicolumn{1}{c}{* Frozen at best fit value.}
	\end{tabular}}
\end{table}

\subsection{Spectro-Polarimetric Properties}
\label{sec:sppol}
\quad

%
%
%
%

\begin{table}
	\centering
	\caption{Polarization parameters obtained using the \texttt{PCUBE} algorithm (for all 3 DUs combined) in different energy bands, for both IXPE Epochs. The uncertainties are reported at the 1$\sigma$ level.}
	\label{tab:pcubetable}
	\resizebox{0.8\columnwidth}{!}{
	\begin{tabular}{cccc} 

	\multicolumn{4}{c}{Epoch 1} \\		
	\hline
	Parameter & $2-4$ keV & $4-8$ keV & $2-8$ keV\\
	\hline
	Q/I (\%) & 1.50 $\pm$  0.36 & 1.04 $\pm$  0.60 & 1.31 $\pm$  0.36\\
	U/I (\%) & $-3.58 \pm$ 0.36 & 5.42 $\pm$  0.60 & $-4.34 \pm$  0.36\\
	PD (\%) & 3.89 $\pm$ 0.36  & 5.52 $\pm$  0.60 & 4.53 $\pm$  0.36\\
	PA ($^\circ$) & 146 $\pm$ 3 & 140 $\pm$ 3 & 143 $\pm$ 2\\

	\hline
	\multicolumn{4}{c}{Epoch 2} \\
	\hline
	Parameter & $2-4$ keV & $4-8$ keV & $2-8$ keV\\
	\hline
	PD (\%) & 0.51 $\pm$ 0.33  & 1.60 $\pm$  0.58 & 0.83 $\pm$  0.33\\
	MDP99(\%) & 1.01 & 1.78 & 1.00\\
	\hline	
\end{tabular}}
\end{table}

The polarization parameters derived from \texttt{PCUBE} for both IXPE Epochs are shown in Table~\ref{tab:pcubetable}. 
We note that the source shows significant polarization during Epoch 1, with PD of $3.89\pm 0.36$\%, $5.52\pm 0.60$\% and $4.53\pm 0.36$\%, (with a significance of 10.7$\sigma$, 9.1$\sigma$ and 12.6$\sigma$ respectively) for the $2-4$, $4-8$ and $2-8$~keV energy bands respectively. The PA for the respective energy bands are 146$^{\circ}$ $\pm$ 3$^{\circ}$, 140$^{\circ}$ $\pm$ 3$^{\circ}$, and 143$^{\circ}$ $\pm$  2$^{\circ}$. For Epoch~2, we did not find significant polarization, with PD 
below the minimum detectable polarizations (MDP) at the 99\% level in all the energy bands (see Table \ref{tab:pcubetable}).

\begin{figure*}
	\centering
	\begin{multicols}{3}
		\hspace{-1.44cm}\subcaptionbox{}{\includegraphics[width=1.21\linewidth]{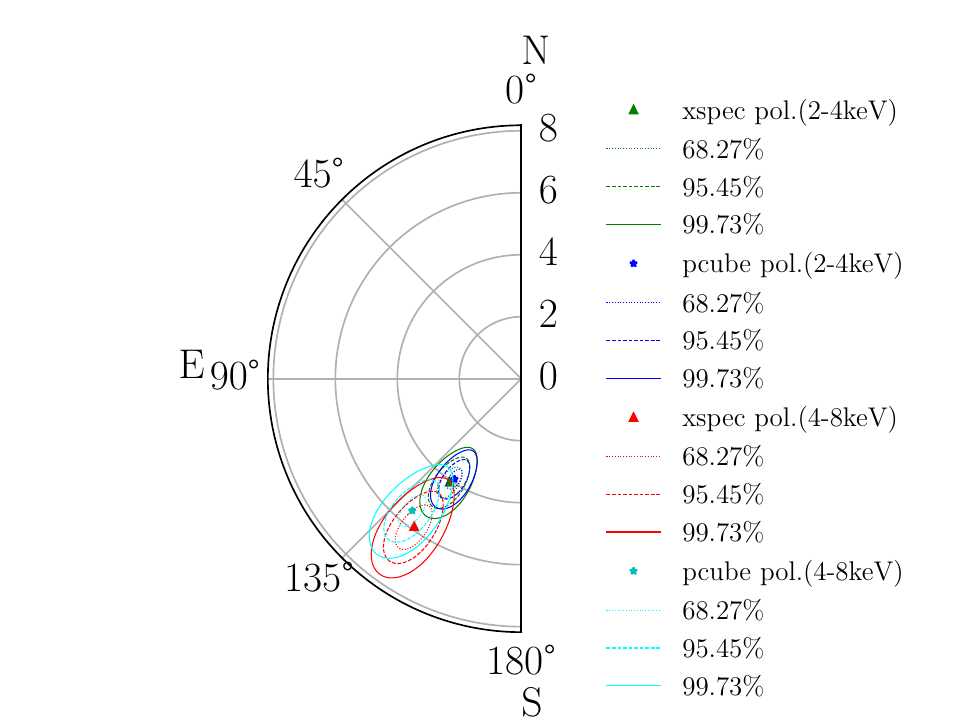}} \par
		\hspace{-1.37cm}\subcaptionbox{}{\includegraphics[width=1.21\linewidth]{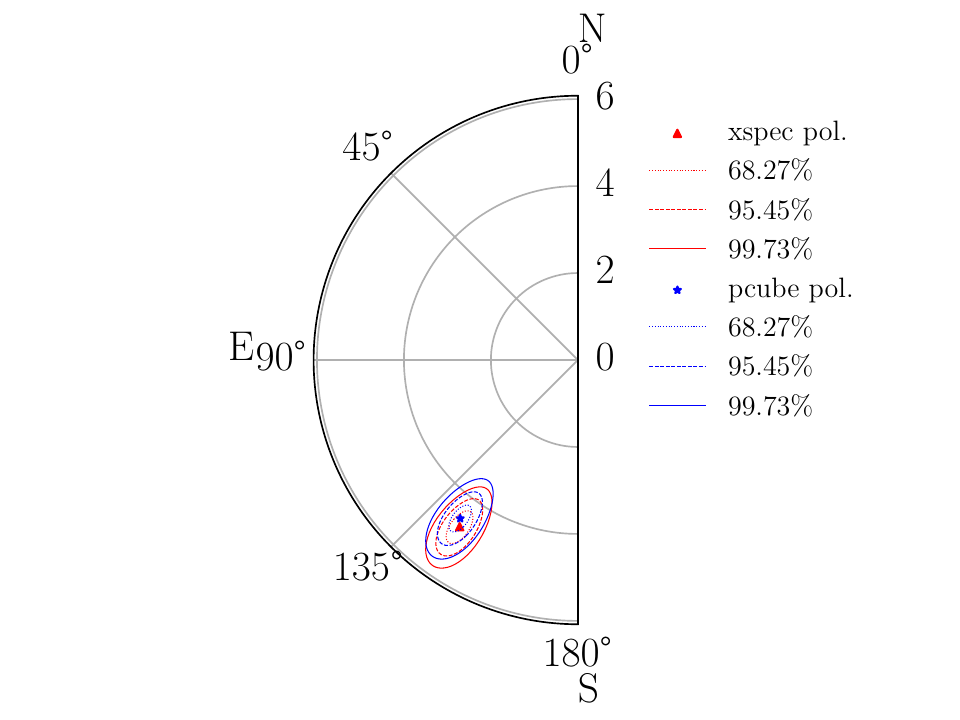}} \par
		\hspace{-1.3cm}\subcaptionbox{}{\includegraphics[width=1.21\linewidth]{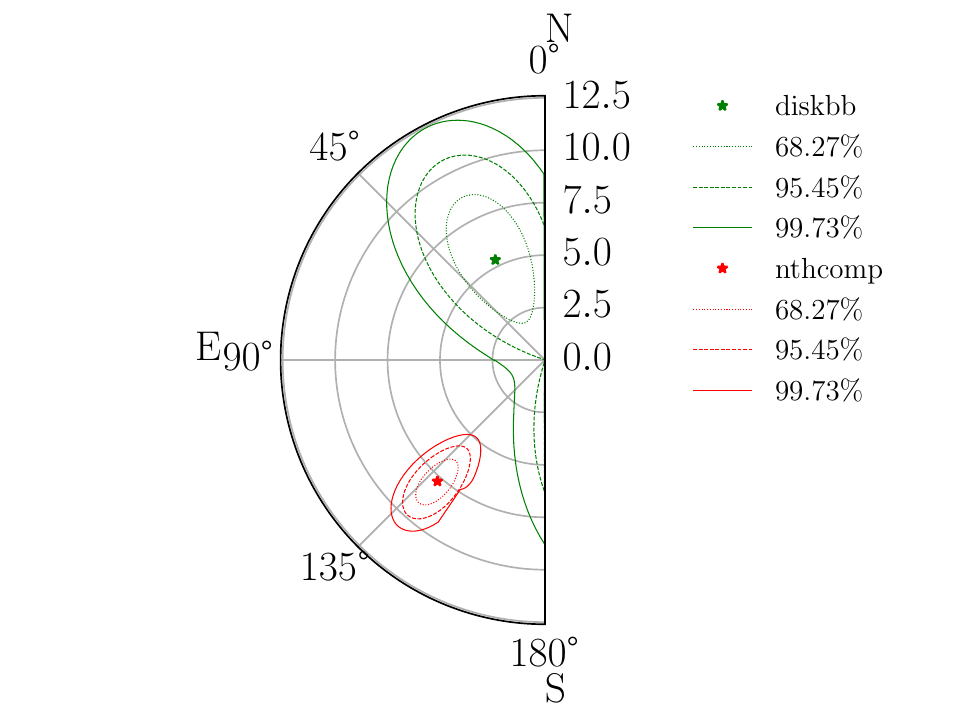}} 		
	\end{multicols}
	\caption{Polarization contours derived from: (a) XSPEC (triangle) and \texttt{PCUBE} (star) in the $2 - 4$ keV (green \& blue) and $4 - 8$ keV (red \& cyan) energy bands; (b) XSPEC (red) and PCUBE (blue) over the $2- 8$ keV energy range; (c) XSPEC using individual \texttt{polconst} for \texttt{diskbb} (green) and \texttt{nthcomp} (red) respectively in the $2-8$ keV energy range. In all three panels, the \textit{IXPE} spectra have been fitted with the best-fit \textit{NICER} parameters using model 1 for Epoch~1, and the 1$\sigma$(dotted), 2$\sigma$(dashed) and 3$\sigma$(solid) contours are shown. See text for details.}
	\label{fig:polar}
\end{figure*}

We also performed a model dependent spectro-polarimetric analysis by simultaneously fitting the source and background spectra for the different Stokes parameters (I, Q \& U) in the different energy bands for Epoch 1. This was carried out for the three models described in Section \ref{sec:specresults}, multiplied by a single \texttt{polconst} (e.g. \texttt{tbabs*polconst*(diskbb+nthComp+gaussian)}).
The parameters were frozen to their best-fit values, leaving only the normalization and polarization parameters to vary freely \citep{fari2023,capi2023}. The results of the spectro-polarimetric fit with model~1 are shown in Table~\ref{tab:xspecfittab}. Figures~\ref{fig:polar}a and \ref{fig:polar}b show the 1$\sigma$, 2$\sigma$, and 3$\sigma$ contours of this fit in the $2-4$, $4-8$ and $2-8$~keV bands. The polarization parameters and contours, although not fully coincident, are in good agreement with those from \texttt{PCUBE}.

We also carried out the spectro-polarimetric analysis of Epoch 1 data with multiple \texttt{polconst}, one multiplied with each component (e.g. \texttt{tbabs*(polconst*diskbb+polconst*nthComp+gaussian)}. The gaussian component is generally not expected to be polarized and hence, we have not multiplied \texttt{polconst} to it.  For model~1, the polarization of \texttt{nthcomp} was well-constrained with PD of $7.7 \pm 2.5$\% and PA of $139\pm9^\circ$. However, the PD of \texttt{diskbb} was unconstrained with a 3$\sigma$ upper limit of 11.5\%. The fit statistic ($\chi^{2}$/DOF) was found to be 1442/1332. The results for model 1 are shown in Table \ref{tab:xspecfittab} and Figure~\ref{fig:polar}c.

For model~2 with multiple \texttt{polconst}, the fit statistic was at 1220/1330 with \texttt{diskbb} PD = $3.3\pm 1.7$\% \& PA = $153^{\circ}\pm 14^{\circ}$ (at the 3$\sigma$ level). However, the PD of the \texttt{bbodyrad} could not be constrained. For model~3, the PD was unconstrained for both \texttt{bbodyrad} and \texttt{nthcomp} components, and the fit statistic was slightly worse at 1661/1330. 
\begin{table}
	\caption{Polarization parameters derived from the spectro-polarimetric fitting of the Stokes spectra of Epoch 1 in different energy bands, using model 1 (top) and model 2 (bottom). Uncertainties and upper limit are quoted at the 3$\sigma$ level. See text for details.}
	\centering
	\label{tab:xspecfittab}
	\resizebox{1\columnwidth}{!}{
		\begin{tabular}{ccccccc}
			
			\multicolumn{6}{c}{Model 1} \\
			\hline
			\multirow{2}{*}{Parameter} & \multicolumn{3}{c}{Single \texttt{polconst}} &  \multicolumn{2}{c}{Multiple \texttt{polconst}$^{*}$ }\\
			\cline{2-6}
			& 2-4 keV    & 4-8 keV   & 2-8 keV  & \texttt{diskbb}            & \texttt{nthcomp} \\
			\hline 
			PD  (\%)                     &  4.05$\pm$1.13       & 5.89$\pm$1.63       & 4.70$\pm$0.93       & $<11.5$        & 7.72$\pm$2.47             \\
			PA   ($^\circ$)                   &    145$\pm$8     & 144$\pm$8      &    145$\pm$5    &   unconstrained             &   139$\pm$9       \\
			$\chi^{2}$/DOF                   & 360/434        & 920/884   &   1442/1334     & \multicolumn{2}{c}{1423/1332}        \\ 
			\hline
\multicolumn{6}{c}{Model 2} \\
\hline
\multirow{2}{*}{Parameter} & \multicolumn{3}{c}{Single \texttt{polconst}} &  \multicolumn{2}{c}{Multiple \texttt{polconst}$^{*}$ }\\
\cline{2-6}
& 2-4 keV    & 4-8 keV   & 2-8 keV  & \texttt{bbodyrad}            & \texttt{diskbb} \\
\hline 
PD  (\%)  &  $4.05 \pm 1.11$     & $5.88 \pm 1.62$       & $4.69\pm 0.92$     & unconstrained        &  $3.33 \pm 1.73$           \\
PA   ($^\circ$)                   &    $145 \pm 8$    & $144 \pm 8$    &   $144 \pm 6$    &   unconstrained            &   153$\pm$14       \\
$\chi^{2}$/DOF                   & 364/434       & 822/884  &   1231/1334    & \multicolumn{2}{c}{1220/1330}        \\ 
\hline    
    \multicolumn{6}{l}{Fitted in the $2-8$ keV range}        
		\end{tabular}
	}
\end{table}

\section{Discussion}
\label{sec:dis}
\quad

Though detailed investigations of spectral and temporal properties
of bright LMXBs have been carried out in the past decades, radiative
processes and nature of accretion flow in these systems still remain
uncertain. Hence, polarization measurements can play a crucial role
in unfolding the nature of accretion flow, emission mechanisms in
these sources. 

 In this work, we report the significant detection of X-ray polarization for IXPE Epoch 1 data as detailed in Section \ref{sec:results}. 
We also carried out model dependent analysis of spectro-polarimetric data using three different models, which provide good fit to the  \textit{IXPE} Epoch 1 data. However, model~3 does not constrain the PD of the individual components. Hence, this model could be excluded from consideration. Model~2 gives low value of PD ($3.3 \pm 1.7$\%) for the disc component and that of the black body component could not be constrained. The simulations carried out by \cite{sch2009} suggest that for direct emission from the accretion disc in the thermal state, the PD decreases with energy and the PA remains constant (see Figure~2 in the cited paper).  However, in our case, the PD increases with energy (Table \ref{tab:xspecfittab}), suggesting that a scenario where polarization arises from a `dominant' accretion disc may not represent the emission spectrum of the source. It should be noted that the best-fit inner disc and blackbody temperatures are unusually high for model 2.

\par The application of model~1 to spectro-polarimetric data gives PD=$7.7 \pm 2.5$\% and PA=$139\pm9^\circ$ for the Comptonized component. The polarization fraction for disc component could not be constrained (see Table~\ref{tab:xspecfittab}). The spectral parameters obtained by \textit{NICER} analysis suggests that during Epoch~1, the source was in the soft state. The presence of iron line with equivalent width $<$ 30~eV, suggests signature of weak reflection. Hence, the orbital inclination of the system is probably close to 75$^\circ$ \citep{lin2009}. 


\par The simulations carried out by \cite{gnar2022} for NS-LMXBs predict that the PD in these sources is $<5$\%. However, the observed PD of the Comptonized component is higher than this. There may be two reasons for this discrepancy. Firstly, their simulations do not consider reflection from the disc which may have resulted in a lower PD. 
Secondly, the narrow energy band of \textit{IXPE} does not allow for detailed spectro-polarimetric modeling of the reflection component. Hence, modeling the data in \textit{XSPEC} with a disc and Comptonized component may result in the polarization due to reflection to be attributed to that due to Comptonization \citep{fari2023}. The spectro-polarimetric analysis shows that the best-fit model consists of emission from a MCD and a Comptonized corona. In this scenario, the Comptonized component is likely to originate from a spreading layer or a boundary layer above the NS surface.

\par During IXPE Epoch~2, we do not detect significant polarization from the source. To investigate this, we modelled the NuSTAR spectrum overlapping with the IXPE observation, using all three models. Using model~1, we find that the corona has become hotter from $kT_{e}=$2.3 to $kT_{e}=$4.5 keV and its optical depth has reduced from $\tau=$11 to $\tau=$2.6. In the other scenario (model~2), the inner disc temperature and blackbody temperature remain nearly same across the two epochs. However, the disc component became stronger and the harder blackbody component became weaker during Epoch~2. 
\par We note that for model 1, the flux of the Comptonized component has weakened compared to Epoch~1 by a factor of $\sim$ 2.9, whereas that of the disc component has increased by a factor of $\sim$ 7.4. As suggested by previous observations of bright LMXBs and present spectro-polarimetric analysis of Epoch~1 data, the disc emission is weakly or probably not polarized \citep{fari2023,rwicha2023}.  Further, a weak corona is not expected to  produce high polarization. Hence, the polarization from the source during Epoch~2 may have fallen below the polarization sensitivity of IXPE.

The same \textit{IXPE} observations of \src were analyzed in parallel to this work by \cite{coc2023}. We note that the authors of the publication have used only \textit{IXPE} data, whereas we have used quasi-simultaneous \textit{NICER} and \textit{NuSTAR} data for the spectral fitting. \cite{coc2023} were admittedly unable to model the \textit{IXPE} spectra (due to its narrow energy band) with a Comptonized emission component which is a better way to describe the wide band spectrum of the source. We have used a Comptonized component in model 1 and model 3 for our analysis. We also note that the moderate energy resolution of the \textit{IXPE} GPD cannot resolve the Gaussian emission line seen in the \textit{NICER} and \textit{NuSTAR} spectra and is therefore not considered in their spectral fitting.

Although our polarimetric results are consistent with that reported in \cite{coc2023}, there is a discrepancy in the spectral fits. We attribute this to the fact that the authors have done the spectral fitting in the narrow 2-8 keV energy band of \textit{IXPE}, resulting in higher $N_{H}$, and lower blackbody and inner disc temperatures. To test this, we performed spectral analysis using model 2 (without the Gaussian component) for the \textit{NICER} and \textit{NuSTAR} data in the 2-8 keV and 3-8 keV band, respectively, by fixing $N_{H}$ as reported in \cite{coc2023}. This exercise resulted in spectral fit parameters similar to theirs. 
We note that spurious line edges not considered in the \textit{NICER} RMF and ARF could have slightly distorted the spectrum and also contributed to the discrepancy in the spectral results. However, the \textit{NuSTAR} data also gives similar values of inner disc temperature and blackbody temperature suggesting that the discrepancy cannot be attributed only to the artifacts in the \textit{NICER} response.
We also found that the spectral fit parameters derived using only \textit{IXPE} data are unable to explain the higher energy spectra of \textit{NuSTAR}
 (> 10 keV). However, our spectro-polarimetric results as well as those of \cite{coc2023}  can be explained by Comptonized emission, probably originating from a spreading-layer/boundary-layer above the neutron star surface. 

A radio jet like emission in the north-south direction had been
observed in this source \citep{fen2007}. However, recent MeerKAT and
ATCA observations of its new outburst suggest that
the large-scale jet are probably background radio galaxies
\citep{gase2022}. Hence, we cannot
comment about the orientation of PA of the polarized X-ray signal with
respect to the symmetry axis.


\section*{Acknowledgments}
\quad

The authors thank the anonymous reviewer for providing valuable suggestions to improve the quality of the manuscript. The authors also thank GH, SAG; DD, PDMSA and Director, URSC for encouragement and continuous support to carry out this research.

This research used data products provided by the \textit{IXPE} team
(MSFC, SSDC, INAF, and INFN) and software tools distributed by the High-Energy Astrophysics Science Archive
Research Center (HEASARC), at NASA Goddard Space Flight
Center (GSFC).

\section*{Data Availability}
\quad

 \textit{IXPE}, \textit{NICER} and \textit{NuSTAR} data underlying this work are available at High Energy Astrophysics Science Archive Research Center
 (HEASARC) facility, located at NASA-Goddard Space Flight Center. 



\bibliographystyle{mnras}
\bibliography{references} 



%
%
%


\bsp	
\label{lastpage}
\end{document}